\documentstyle[amsmath,12pt]{article}

\begin{document}

\begin{titlepage}

\begin{center}

{\Large \bf
The Batalin-Vilkovisky Field-Antifield Action for Systems with
First-Class Constraints}
\vskip .6in

Domingo J. Louis-Martinez
\vskip .2in

Science One Program and\\ 
Department of Physics and Astronomy,\\ University of British
Columbia\\Vancouver, Canada

\end{center}
 
\vskip 3cm
 
\begin{abstract}
The Batalin-Vilkovisky field-antifield action for systems with first-class
constraints is given explicitly in terms of the canonical hamiltonian, 
the hamiltonian constraints and the first-order hamiltonian gauge
structure functions. It is shown that this
action does not depend on the hamiltonian gauge structure functions of
higher orders. 
A method for finding the lagrangian gauge structure tensors
of all orders is presented. It is proven that the lagrangian gauge
structure tensors do not depend on the hamiltonian gauge structure
functions of second- or higher-orders.  
\end{abstract} 
\end{titlepage}

The Batalin-Vilkovisky (BV) quantization approach 
\cite{batalin1, batalin2, gomis, gitman} is a 
reliable quantization scheme for theories with open gauge algebras like 
supergravity \cite{freedman, kallosh, dewit}. 
In the lagrangian formalism, an open gauge algebra is characterized by a
set of gauge structure tensors \cite{dewit, batalin1, batalin2, gomis}. 
These tensors can be obtained from the BV field-antifield action
\cite{batalin2, gomis}. 
In several works
\cite{siegel, henneaux1, batlle1, grigoryan, razumov, dejonghe, batalin3}
it has been shown that the BV field-antifield action is determined by the
BRST extended Hamiltonian and the BRST charge.  In order to determine
the BRST charge, one requires knowledge of the hamiltonian gauge structure
functions of all orders \cite{henneaux}. Therefore, one may conclude that 
the lagrangian
gauge structure tensors of higher orders must depend on the higher order
hamiltonian gauge structure functions. In this paper we will show that
this is not the case.

The purpose of the present paper is to prove that {\it all} the lagrangian
gauge structure tensors are completely determined by 
the canonical Hamiltonian, the hamiltonian first-class constraints and the 
hamiltonian first-order gauge structure functions. In order to accomplish 
this, we will show that the generating functional of the lagrangian 
gauge structure tensors, the BV  field-antifield action, can be written 
as a function of these hamiltonian quantities alone and 
does not depend on the hamiltonian gauge structure tensors of higher
orders. We will also present an algorithm that will allow us to find all
the lagrangian gauge structure tensors explicitly.

The BV field-antifield action $S$ is a proper solution of the 
classical master equation \cite{batalin2, gomis}:

\begin{equation}
\left(S, S\right) \equiv 0
\label{3.2}
\end{equation}

with the boundary conditions:

\begin{equation}
\left. S\right|_{q^{*}=0, {\cal C}^{*}=0} = S_{0}
\label{3.2a}
\end{equation}

\begin{equation}
\left. \frac{\delta_{l}\delta_{r} S}{\delta q^{*}_{i} \delta
{\cal C}^{\mu}}
\right|_{q^{*}=0, {\cal C}^{*}=0} = R_{\mu}^{i}
\label{3.2b}
\end{equation}

The bracket in (\ref{3.2}) is the BV antibracket \cite{batalin2}.

The quantities ${\cal C}^{\mu} (\mu = 1,2,...,m)$ are ghost fields. 
The variables
$q^{*}_{i} (i= 1,2,...,n)$ and ${\cal C}^{*}_{\mu} (\mu = 1,2,...,m)$ are
the antifields \cite{batalin2}. 

The zeroth-order lagrangian gauge structure function $S_{0}$ is the action
functional of the physical theory:

\begin{equation}
S_{0} = \int dt L_{0}
\label{3.1a}
\end{equation}

The lagrangian first-order gauge structure functions $R_{\mu}^{i}$ are the
generators of the lagrangian gauge transformations.

The BV field-antifield action can be written in the following form:

\begin{equation}
S = \int dt L
\label{3.3}
\end{equation}

\noindent where \cite{batalin2, gomis}:

\begin{eqnarray}
L & = & L_{0} + q^{*}_{i} R_{\alpha}^{i}{\cal C}^{\alpha} +
{1 \over 2} {\cal C}^{*}_{\delta} T_{\alpha\beta}^{\delta} {\cal
C}^{\beta}
{\cal C}^{\alpha} - {1 \over 4} q^{*}_{i} q^{*}_{j} E_{\alpha\beta}^{ji}
{\cal C}^{\beta}{\cal C}^{\alpha} - {1 \over 2} {\cal C}^{*}_{\delta}
q^{*}_{i}
D_{\alpha\beta\gamma}^{i\delta} {\cal C}^{\gamma} {\cal C}^{\beta} 
{\cal C}^{\alpha} +\nonumber\\
& & + {1 \over 12} q^{*}_{i} q^{*}_{j} q^{*}_{k} 
M_{\alpha\beta\gamma}^{kji} {\cal C}^{\gamma}{\cal C}^{\beta}
{\cal C}^{\alpha} + ...
\label{3.4}
\end{eqnarray}

The tensors $T_{\alpha\beta}^{\eta}$ and $E_{\alpha\beta}^{ij}$ are the 
second-order gauge structure functions of the gauge algebra in the
lagrangian formalism. 
The generators of lagrangian gauge transformations $R_{\mu}^{i}$ are said 
to form an open gauge algebra if
$E_{\alpha\beta}^{ij} \neq 0$ \cite{dewit, batalin1, batalin2,
gomis}. If $E_{\alpha\beta}^{ij} = 0$, the gauge algebra is said to be
closed.

The existence of the higher order lagrangian gauge structure functions
($D_{\alpha\beta\gamma}^{i\rho}, M_{\alpha\beta\gamma}^{ijk}$, etc) has 
been proven using an axiomatic approach in \cite{batalin2}. 
These functions were constructed explicitly in \cite{algebra}.

The BV field-antifield action is the generating functional of the
structure tensors of the gauge algebra in the lagrangian formalism
\cite{batalin2, gomis}.

Let us consider a
system with only primary first-class irreducible hamiltonian constraints 
$G_{\mu} (\mu = 1,2,...,m)$, such that \cite{dirac, fradkin}:

\begin{equation}
\{G_{\alpha}, G_{\beta}\} \equiv C_{\alpha\beta}^{\eta} G_{\eta}
\label{3.5}
\end{equation}

\begin{equation}
\{H_{0}, G_{\mu}\} \equiv V_{\mu}^{\eta} G_{\eta}
\label{3.6}
\end{equation}

\begin{equation}
\{C_{\alpha\beta}^{\eta}, G_{\gamma}\} 
+ \{C_{\beta\gamma}^{\eta}, G_{\alpha}\} 
+ \{C_{\gamma\alpha}^{\eta}, G_{\beta}\} 
- C_{\alpha\beta}^{\delta} C_{\gamma\delta}^{\eta} 
- C_{\beta\gamma}^{\delta} C_{\alpha\delta}^{\eta} 
- C_{\gamma\alpha}^{\delta} C_{\beta\delta}^{\eta} 
\equiv J_{\alpha\beta\gamma}^{\eta\sigma} G_{\sigma}
\label{3.A}
\end{equation}

The function $H_{0}$ is the canonical Hamiltonian. The tensors 
$C_{\alpha\beta}^{\eta}$ and $J_{\alpha\beta\gamma}^{\eta\sigma}$
are the hamiltonian gauge structure functions of first- and second-order
respectively.

Let ${\rm FL}^*$ be the pullback application \cite{batlle, chaichian}
from the momentum phase space into the velocity phase space defined by the
relations: 

\begin{equation}
{\rm FL}^* p_{i} \equiv \frac{\partial
L_{0}}{\partial\dot{q}^{i}}(q,\dot{q})
\label{3.B}
\end{equation}

Notice that the generators of the lagrangian gauge transformations 
$R_{\mu}^{i}$
can be written in terms of the hamiltonian constraints as follows 
\cite{dirac}:

\begin{equation}
R_{\mu}^{i} = {\rm FL}^* \frac{\partial G_{\mu}}{\partial p_{i}}
\label{3.7b}
\end{equation}

Let us consider the extended momentum phase space with points 
$(q^{i}, {\cal P}_{i}, {\cal C}^{\alpha}, \pi_{\alpha}, q^{*}_{j}, p^{*j},
{\cal C}^{*}_{\beta}, \pi^{* \beta})$, where ${\cal P}_{i}$ are the
canonical momenta conjugate to $q^{i}$:

\begin{eqnarray}
{\cal P}_{i} & = & p_{i} +  q^{*}_{k} p_{i \alpha}^{k}(q, p) {\cal
C}^{\alpha}
+
{1 \over 2} {\cal C}^{*}_{\delta} p_{i \alpha\beta}^{\delta}(q, p) 
{\cal C}^{\beta} {\cal C}^{\alpha} - {1 \over 4} q^{*}_{k} q^{*}_{l}
p_{i \alpha\beta}^{lk}(q, p)
{\cal C}^{\beta}{\cal C}^{\alpha} \nonumber\\
& &  - {1 \over 2} {\cal C}^{*}_{\delta}
q^{*}_{k}
p_{i \alpha\beta\gamma}^{k \delta}(q, p) {\cal C}^{\gamma} {\cal
C}^{\beta} 
{\cal C}^{\alpha} 
+ {1 \over 12} q^{*}_{m} q^{*}_{l} q^{*}_{k} 
p_{i \alpha\beta\gamma}^{klm}(q, p) {\cal C}^{\gamma}{\cal C}^{\beta}
{\cal C}^{\alpha} + ...
\label{3.9}
\end{eqnarray}

\noindent The canonical momentum variables $\pi_{\alpha}$, $p^{* j}$ and
$\pi^{* \beta}$ are conjugate to the coordinates ${\cal C}^{\alpha}$,
$q^{*}_{j}$ and 
${\cal C}^{*}_{\beta}$ respectively.

We define the pullback application ${\cal FL}^*$ as follows:

\begin{eqnarray}
{\cal FL}^* {\cal P}_{i} & \equiv & 
{\rm FL}^* p_{i} +  q^{*}_{k}{\rm FL}^* p_{i \alpha}^{k} 
{\cal C}^{\alpha} +
{1 \over 2} {\cal C}^{*}_{\delta}{\rm FL}^* p_{i \alpha\beta}^{\delta} 
{\cal C}^{\beta} {\cal C}^{\alpha} - {1 \over 4} q^{*}_{k} q^{*}_{l}
{\rm FL}^* p_{i \alpha\beta}^{lk}
{\cal C}^{\beta}{\cal C}^{\alpha} \nonumber\\ 
& &  - {1 \over 2} {\cal C}^{*}_{\delta}
q^{*}_{k}
{\rm FL}^* p_{i \alpha\beta\gamma}^{k \delta} {\cal C}^{\gamma} {\cal
C}^{\beta} 
{\cal C}^{\alpha} 
+ {1 \over 12} q^{*}_{m} q^{*}_{l} q^{*}_{k} 
{\rm FL}^* p_{i \alpha\beta\gamma}^{klm} {\cal C}^{\gamma}{\cal
C}^{\beta} {\cal C}^{\alpha} + ...
\label{3.10}
\end{eqnarray}

\begin{equation}
{\cal FL}^* \pi_{\alpha} \equiv 0
\label{3.11}
\end{equation}

\begin{equation}
{\cal FL}^* p^{* j} \equiv 0
\label{3.12}
\end{equation}

\begin{equation}
{\cal FL}^* \pi^{* \beta} \equiv 0
\label{3.13}
\end{equation}

The functions ${\rm FL}^* p_{i}$, ${\rm FL}^* p_{i\alpha}^{k}$, 
${\rm FL}^* p_{i\alpha\beta}^{\delta}$, 
${\rm FL}^* p_{i\alpha\beta}^{lk}$, 
${\rm FL}^* p_{i\alpha\beta\gamma}^{k\delta}$, 
${\rm FL}^* p_{i\alpha\beta\gamma}^{klm}$, etc are determined by
the equations:

\begin{equation}
{\cal FL}^* {\cal P}_{i} \equiv \frac{\partial L}{\partial \dot{q}^{i}}
\label{3.C}
\end{equation}

\begin{equation}
{\cal FL}^* \frac{\partial H}{\partial {\cal P}_{i}} +
\Lambda^{\mu} {\cal FL}^* \frac{\partial G_{\mu}}{\partial {\cal P}_{i}}
\equiv \dot{q}^{i}
\label{3.14}
\end{equation}

\begin{equation}
{\cal FL}^* G_{\mu} \equiv 0
\label{3.15}
\end{equation}

We claim that the BV Lagrangian $L$ can be written in terms of the 
canonical Hamiltonian $H_{0}$, the hamiltonian first-class constraints 
$G_{\mu}$ and the 
hamiltonian first-order gauge structure funtions $C_{\mu\nu}^{\eta}$ as 
follows:

\begin{equation}
L = \dot{q}^{i} {\cal FL}^* {\cal P}_{i} - {\cal FL}^* H_{0}
+ q^{*}_{k} {\cal FL}^* \frac{\partial G_{\alpha}}{\partial {\cal P}_{k}}
{\cal C}^{\alpha} + {1 \over 2} {\cal C}^{*}_{\delta} {\cal FL}^*
C_{\alpha\beta}^{\delta} {\cal C}^{\beta} {\cal C}^{\alpha}
\label{3.20}
\end{equation}

\noindent The hamiltonian $H$ will be given by the expression:

\begin{equation}
H = H_{0}\left(q, {\cal P}\right) - q^{*}_{k} \frac{\partial
G_{\alpha}}{\partial
{\cal P}_{k}}\left(q, {\cal P}\right) {\cal C}^{\alpha}
- {1 \over 2} {\cal C}^{*}_{\delta} 
C_{\alpha\beta}^{\delta}\left(q, {\cal P}\right) {\cal C}^{\beta} 
{\cal C}^{\alpha} 
\label{3.16}
\end{equation}

From (\ref{3.10}) it follows that for any analytic function $K =
K\left(q, {\cal P}\right)$ we can write:

\begin{eqnarray}
{\cal FL}^* K & \equiv & K\left(q, {\cal FL}^* {\cal P}\right)\nonumber\\
& \equiv & {\rm FL}^* K +  q^{*}_{k}{\rm FL}^* 
\left(\frac{\partial K}{\partial p_{i}}p_{i \alpha}^{k}\right) 
{\cal C}^{\alpha} +
{1 \over 2} {\cal C}^{*}_{\delta}{\rm FL}^* 
\left(\frac{\partial K}{\partial p_{i}} p_{i \alpha\beta}^{\delta}\right) 
{\cal C}^{\beta} {\cal C}^{\alpha} \nonumber\\
& & - {1 \over 4} q^{*}_{k} q^{*}_{l}
{\rm FL}^* \left( \frac{\partial K}{\partial p_{i}} 
p_{i\alpha\beta}^{lk} - \frac{\partial^{2} K}{\partial p_{i}
\partial p_{j}} \left(p_{i\alpha}^{k} p_{j\beta}^{l} -
p_{i\beta}^{k} p_{j\alpha}^{l}\right)\right) {\cal C}^{\beta}{\cal
C}^{\alpha} \nonumber\\ 
& &  - {1 \over 2} {\cal C}^{*}_{\delta} q^{*}_{k}
{\rm FL}^* \left(\frac{\partial K}{\partial p_{i}} 
p_{i\alpha\beta\gamma}^{k \delta} - {1 \over 3} \frac{\partial^{2}
K}{\partial p_{i} \partial p_{j}} \left(p_{i\alpha}^{k}
p_{j\beta\gamma}^{\delta} + p_{i\beta}^{k}
p_{j\gamma\alpha}^{\delta} + p_{i\gamma}^{k}
p_{j\alpha\beta}^{\delta}\right) \right) {\cal C}^{\gamma} {\cal
C}^{\beta} {\cal C}^{\alpha} \nonumber\\
& & + {1 \over 12} q^{*}_{n} q^{*}_{m} q^{*}_{l} 
{\rm FL}^* \left(\frac{\partial K}{\partial p_{i}} 
p_{i\alpha\beta\gamma}^{lmn} - \frac{\partial^{2} K}{\partial
p_{i} \partial p_{j}} \left(p_{i\alpha}^{n} p_{j\beta\gamma}^{lm} 
+ p_{i\beta}^{n} p_{j\gamma\alpha}^{lm} 
+ p_{i\gamma}^{n} p_{j\alpha\beta}^{lm}\right)\right.\nonumber\\
& &\hspace{3.4cm} \left. + 2 \frac{\partial^{3} K}{\partial p_{i} \partial
p_{j} \partial
p_{k}}
p_{i\alpha}^{n} p_{j\beta}^{m} p_{k\gamma}^{l}\right)
{\cal C}^{\gamma}{\cal C}^{\beta} {\cal C}^{\alpha} + ...
\label{3.18}
\end{eqnarray}

\noindent The Lagrange multipliers $\Lambda^{\mu}$ can also be 
written as:

\begin{equation}
\Lambda^{\mu} = \lambda^{\mu}(q, \dot{q}) 
+ q^{*}_{k} \lambda^{\mu k}_{\alpha}(q, \dot{q}) {\cal C}^{\alpha}
+ {1 \over 2} {\cal C}^{*}_{\delta} 
\lambda^{\mu\delta}_{\alpha\beta}(q, \dot{q}) {\cal C}^{\beta} 
{\cal C}^{\alpha} + ...
\label{3.17}
\end{equation}

Let us consider the action:

\begin{equation}
S = \int dt L
\label{3.19}
\end{equation}

\noindent where $L$ is given by the expression (\ref{3.20}).

We will prove that the action $S$ (\ref{3.19},\ref{3.20}) is a proper
solution
of the classical master equation (\ref{3.2}) with the boundary conditions
(\ref{3.2a}) and (\ref{3.2b}).

The antibracket in (\ref{3.2}) can be written as \cite{batalin2, gomis}:

\begin{equation}
\left(S, S\right) \equiv 2 \int dt 
\left(\frac{\delta_{r} S}{\delta q^{i}} \frac{\delta_{l} S}{\delta
q^{*}_{i}} + \frac{\delta_{r} S}{\delta {\cal C}^{\alpha}}
\frac{\delta_{l} S}{\delta {\cal C}^{*}_{\alpha}}\right)
\label{3.21}
\end{equation}

From (\ref{3.20}) and (\ref{3.16}) we find that the first functional
derivatives of the action $S$ can be written in the following form:

\begin{equation}
\frac{\delta_{r} S}{\delta q^{i}}  = 
{\cal FL}^* {\cal P}_{i} \frac{d}{dt} 
- {\cal FL}^* \frac{\partial H}{\partial q^{i}} 
+ \left(
\dot{q}^{j} - {\cal FL}^{*} \frac{\partial H}{\partial {\cal P}_{j}}
\right)
\left(\frac{\partial}{\partial q^{i}}\left({\cal FL}^*
{\cal P}_{j}\right) 
+ \frac{\partial}{\partial \dot{q}^{i}}
\left({\cal FL}^* {\cal P}_{j}\right) \frac{d}{dt}\right)
\label{3.22}
\end{equation}

\begin{equation}
\frac{\delta_{l} S}{\delta q^{*}_{i}}  = 
{\cal FL}^* \frac{\partial G_{\alpha}}{\partial {\cal P}_{i}} {\cal
C}^{\alpha}
+ \frac{\partial_{l}}{\partial q^{*}_{i}} \left({\cal FL}^*
{\cal P}_{j}\right) \left(
\dot{q}^{j} - {\cal FL}^{*} \frac{\partial H}{\partial {\cal P}_{j}}
\right)
\label{3.23}
\end{equation}

\begin{equation}
\frac{\delta_{r} S}{\delta {\cal C}^{\alpha}}  = 
q^{*}_{k} {\cal FL}^* \frac{\partial G_{\alpha}}{\partial {\cal P}_{k}}
+ {\cal C}^{*}_{\delta} {\cal FL}^* C_{\alpha\beta}^{\delta} {\cal
C}^{\beta} 
+ \left(
\dot{q}^{j} - {\cal FL}^{*} \frac{\partial H}{\partial {\cal P}_{j}}\right)
\frac{\partial_{r}}{\partial {\cal C}^{\alpha}} \left({\cal 
FL}^* {\cal P}_{j}\right)
\label{3.24}
\end{equation}

\begin{equation}
\frac{\delta_{l} S}{\delta {\cal C}^{*}_{\alpha}}  = 
{1 \over 2} {\cal FL}^* C_{\beta\gamma}^{\alpha} {\cal C}^{\gamma}
{\cal C}^{\beta} 
+ \frac{\partial_{l}}{\partial {\cal C}^{*}_{\alpha}} \left({\cal 
FL}^* {\cal P}_{j}\right)\left(
\dot{q}^{j} - {\cal FL}^{*} \frac{\partial H}{\partial {\cal P}_{j}}\right)
\label{3.25}
\end{equation}

On the other hand, from (\ref{3.15}) we find the following identities:

\begin{equation}
\frac{\partial}{\partial q^{i}} \left({\cal FL}^* G_{\mu}\right)
\equiv {\cal FL}^* \frac{\partial G_{\mu}}{\partial q^{i}} 
+ \frac{\partial}{\partial q^{i}} \left({\cal FL}^* {\cal P}_{j}\right)
{\cal FL}^* \frac{\partial G_{\mu}}{\partial {\cal P}_{j}}
\equiv 0
\label{3.27}
\end{equation}

\begin{equation}
\frac{\partial}{\partial \dot{q}^{i}} \left({\cal FL}^* G_{\mu}\right)
\equiv \frac{\partial}{\partial \dot{q}^{i}} \left({\cal FL}^*
{\cal P}_{j}\right) {\cal FL}^* \frac{\partial G_{\mu}}{\partial {\cal
P}_{j}} \equiv 0
\label{3.28}
\end{equation}

\begin{equation}
\frac{\partial_{l}}{\partial q^{*}_{i}} \left({\cal FL}^* G_{\mu}\right)
\equiv \frac{\partial_{l}}{\partial q^{*}_{i}} \left({\cal FL}^*
{\cal P}_{j}\right) {\cal FL}^* \frac{\partial G_{\mu}}{\partial {\cal
P}_{j}} \equiv 0
\label{3.29}
\end{equation}

\begin{equation}
\frac{\partial_{r}}{\partial {\cal C}^{\alpha}} \left({\cal FL}^*
G_{\mu}\right)
\equiv {\cal FL}^* \frac{\partial G_{\mu}}{\partial {\cal P}_{j}} 
\frac{\partial_{r}}{\partial {\cal C}^{\alpha}} \left({\cal FL}^*
{\cal P}_{j}\right) \equiv 0
\label{3.30}
\end{equation}

\begin{equation}
\frac{\partial_{l}}{\partial {\cal C}^{*}_{\alpha}} \left({\cal FL}^*
G_{\mu}\right)
\equiv \frac{\partial_{l}}{\partial {\cal C}^{*}_{\alpha}} \left({\cal
FL}^*
{\cal P}_{j}\right) {\cal FL}^* \frac{\partial G_{\mu}}{\partial {\cal
P}_{j}} \equiv 0
\label{3.31}
\end{equation}

Substituting (\ref{3.14}) and (\ref{3.27}-\ref{3.31}) into
(\ref{3.22}-\ref{3.25}) we finally obtain:

\begin{equation}
\frac{\delta_{r} S}{\delta q^{i}}  = 
{\cal FL}^* {\cal P}_{i} \frac{d}{dt} 
- {\cal FL}^* \frac{\partial }{\partial q^{i}} 
- \Lambda^{\mu} {\cal FL}^* \frac{\partial G_{\mu}}{\partial q^{i}}
\label{3.32}
\end{equation}

\begin{equation}
\frac{\delta_{l} S}{\delta q^{*}_{i}}  = 
{\cal FL}^* \frac{\partial G_{\alpha}}{\partial {\cal P}_{i}} {\cal
C}^{\alpha}
\label{3.33}
\end{equation}

\begin{equation}
\frac{\delta_{r} S}{\delta {\cal C}^{\alpha}}  = 
q^{*}_{k} {\cal FL}^* \frac{\partial G_{\alpha}}{\partial {\cal P}_{k}}
+ {\cal C}^{*}_{\delta} {\cal FL}^* C_{\alpha\beta}^{\delta} {\cal
C}^{\beta} 
\label{3.34}
\end{equation}

\begin{equation}
\frac{\delta_{l} S}{\delta {\cal C}^{*}_{\alpha}}  = 
{1 \over 2} {\cal FL}^* C_{\beta\gamma}^{\alpha} {\cal C}^{\gamma}
{\cal C}^{\beta} 
\label{3.35}
\end{equation}

Therefore, we can write the BV antibracket of the action $S$ with itself
as follows:

\begin{eqnarray}
\left(S, S\right) & = &
2 \int dt \left[{\cal FL}^* {\cal P}_{i} \frac{d}{dt}\left({\cal FL}^*
\frac{\partial G_{\alpha}}{\partial {\cal P}_{i}} {\cal C}^{\alpha}\right)
+ {1 \over 2} q^{*}_{j} {\cal FL}^* \frac{\partial G_{\mu}}{\partial
{\cal P}_{j}} {\cal FL}^* C_{\alpha\beta}^{\mu} {\cal C}^{\beta} {\cal
C}^{\alpha} \right.\nonumber\\
& & - \left(
{\cal FL}^* \frac{\partial H}{\partial q^{i}} 
+ \Lambda^{\eta} {\cal FL}^* \frac{\partial
G_{\eta}}{\partial q^{i}}\right) {\cal FL}^* \frac{\partial
G_{\mu}}{\partial {\cal P}_{i}} {\cal C}^{\mu}\nonumber\\
& & \left.- {1 \over 6} {\cal C}^{*}_{\delta} {\cal FL}^* \left(
C_{\alpha\beta}^{\eta} C_{\gamma\eta}^{\delta} 
+ C_{\beta\gamma}^{\eta} C_{\alpha\eta}^{\delta} 
+ C_{\gamma\alpha}^{\eta} C_{\beta\eta}^{\delta} 
\right) {\cal C}^{\gamma} {\cal C}^{\beta} {\cal C}^{\alpha}\right] 
\label{3.36}
\end{eqnarray}

From (\ref{3.36}), integrating by parts and using the relations
(\ref{3.14}, \ref{3.16}) we find that the BV antibracket of the
functional $S$ with itself can be written as:

\begin{eqnarray}
\left(S, S\right) & = &
2 \int dt \left[ - \frac{d}{dt} \left({\cal FL}^* G_{\alpha}\right)
{\cal C}^{\alpha} + {\cal FL}^* \{G_{\alpha}, H_{0}\} {\cal C}^{\alpha}
+ \Lambda^{\mu} {\cal FL}^* \{G_{\alpha}, G_{\mu}\} {\cal
C}^{\alpha}\right. \nonumber\\
& & - {1 \over 2} q^{*}_{k} {\cal FL}^*
\left(\frac{\partial}{\partial {\cal P}_{k}}\{G_{\alpha},
G_{\beta}\} - \frac{\partial G_{\eta}}{\partial {\cal P}_{k}}
C_{\alpha\beta}^{\eta}\right) {\cal C}^{\beta} {\cal
C}^{\alpha}\nonumber\\
& &\left. - {1 \over 6} {\cal C}^{*}_{\delta} {\cal FL}^* \left(
\{G_{\alpha}, C_{\beta\gamma}^{\delta}\} 
+ \{G_{\beta}, C_{\gamma\alpha}^{\delta}\} 
+ \{G_{\gamma}, C_{\alpha\beta}^{\delta}\} 
+ C_{\alpha\beta}^{\eta} C_{\gamma\eta}^{\delta} 
+ C_{\beta\gamma}^{\eta} C_{\alpha\eta}^{\delta} 
+ C_{\gamma\alpha}^{\eta} C_{\beta\eta}^{\delta} 
\right) {\cal C}^{\gamma} {\cal C}^{\beta} {\cal C}^{\alpha}\right]
\nonumber\\
& & + \left. 2 {\cal FL}^* \left({\cal P}_{i} \frac{\partial
G_{\alpha}}{\partial {\cal P}_{i}}\right) {\cal
C}^{\alpha}\right|^{t_f}_{t_i}
\label{3.37}
\end{eqnarray}

Using (\ref{3.5}, \ref{3.6}, \ref{3.A}) and (\ref{3.15}) in (\ref{3.37}),
it immediately follows that (\ref{3.37}) reduces to:

\begin{equation}
\left(S, S\right) =
\left. 2 {\cal FL}^* \left({\cal P}_{i} \frac{\partial
G_{\alpha}}{\partial {\cal P}_{i}}\right) {\cal
C}^{\alpha}\right|^{t_f}_{t_i}
\label{3.38}
\end{equation}

Assuming that ${\cal C}^{\alpha}(t_i) = {\cal C}^{\alpha}(t_f) = 0$ we
finally obtain:

\begin{equation}
\left(S, S\right) = 0 
\label{3.39}
\end{equation}

This proves that $S$ in (\ref{3.19}, \ref{3.20}) satisfies the BV
classical master equation.

From (\ref{3.20}) and (\ref{3.9}) it immediately follows that:

\begin{equation}
\left. L\right|_{q^{*} = 0, {\cal C}^{*} = 0} =
\dot{q}^{i} {\rm FL}^* p_{i} - {\rm FL}^* H_{0} = L_{0}
\label{3.40}
\end{equation}

\noindent and therefore, $S$ in (\ref{3.19}, \ref{3.20}) satisfies the
boundary condition (\ref{3.2a}).

From (\ref{3.33}) and (\ref{3.34}) it follows that:

\begin{equation}
\left. \frac{\delta_{l}\delta_{r} S}{\delta q^{*}_{i} \delta
{\cal C}^{\mu}}
\right|_{q^{*}=0, {\cal C}^{*}=0} = {\cal FL}^* \frac{\partial
G_{\alpha}}{\partial {\cal P}_{k}}
\label{3.41}
\end{equation}

Finally, from (\ref{3.41}), (\ref{3.18}) and (\ref{3.7b}) we obtain:

\begin{equation}
\left. \frac{\delta_{l}\delta_{r} S}{\delta q^{*}_{i} \delta
{\cal C}^{\mu}}
\right|_{q^{*}=0, {\cal C}^{*}=0} = R_{\mu}^{i}
\label{3.42}
\end{equation}

This proves that the action functional $S$ also satisfies the boundary
conditions (\ref{3.2b}).

Therefore, we conclude that the Batalin-Vilkovisky field-antifield
action can be written in the form:

\begin{equation}
S = \int dt \left[\dot{q}^{i} {\cal FL}^* {\cal P}_{i} - {\cal FL}^*
H_{0}
+ q^{*}_{k} {\cal FL}^* \frac{\partial G_{\alpha}}{\partial {\cal P}_{k}}
{\cal C}^{\alpha} + {1 \over 2} {\cal C}^{*}_{\delta} {\cal FL}^*
C_{\alpha\beta}^{\delta} {\cal C}^{\beta} {\cal C}^{\alpha}\right]
\label{3.D}
\end{equation}

As it can be seen from (\ref{3.D}), the BV field-antifield action is
completely determined by the canonical hamiltonian $H_{0}$, the hamiltonian
constraints $G_{\mu}$ and the hamiltonian first-order gauge structure
functions $C_{\mu\nu}^{\eta}$. It does not depend on the hamiltonian
second-order gauge structure functions
$J_{\alpha\beta\gamma}^{\eta\sigma}$ or other higher order hamiltonian
gauge structure functions. Since the BV field-antifield action is the
generating functional of all the lagrangian gauge structure tensors
(\ref{3.4}), we conclude that the lagrangian gauge structure
tensors of all orders  are completely determined by $H_{0}$,
$G_{\mu}$ and $C_{\mu\nu}^{\eta}$.

Notice that formula ({\ref{3.D}}) is valid for the generic case of
systems with {\it open} lagrangian gauge algebras. These systems may 
have nonvanishing lagrangian gauge structure tensors of higher orders.

From (\ref{3.D}) using (\ref{3.10}) and (\ref{3.18}) 
and expanding (\ref{3.14}) and (\ref{3.15}) with the use of (\ref{3.10}),
(\ref{3.18}) and (\ref{3.16}, \ref{3.17}) we obtain  
the lagrangian gauge structure tensors up to fourth order in the
form:

\begin{subequations}
\begin{equation}
S_{0} = \int dt \left[\dot{q}^{i} {\rm FL}^* p_{i} - {\rm FL}^*
H_{0}\right]
\label{3.48a}
\end{equation}

\begin{equation}
R_{\mu}^{i} = {\rm FL}^* \frac{\partial G_{\mu}}{\partial p_{i}}
\label{3.48}
\end{equation}

\begin{equation}
T_{\mu\nu}^{\eta} = {\rm FL}^* C_{\mu\nu}^{\eta}
\label{3.49}
\end{equation}

\begin{equation}
E_{\mu\nu}^{ij}  =  
{\rm FL}^* \left(p_{k \mu}^{i} 
\frac{\partial^{2} G_{\nu}}{\partial p_{k} \partial p_{j}}
- p_{k \nu}^{i} \frac{\partial^{2} G_{\mu}}{\partial p_{k} \partial
p_{j}}\right) 
\label{3.50}
\end{equation}

\begin{equation}
D_{\alpha\beta\gamma}^{i\rho}  = 
- {1 \over 3} {\rm FL}^* \left(p_{k \alpha}^{i}
\frac{\partial C_{\beta\gamma}^{\rho}}{\partial p_{k}}
+ p_{k \beta}^{i}
\frac{\partial C_{\gamma\alpha}^{\rho}}{\partial p_{k}}
+ p_{k \gamma}^{i}
\frac{\partial C_{\alpha\beta}^{\rho}}{\partial p_{k}}\right)
\label{3.51}
\end{equation}

\begin{eqnarray}
M_{\alpha\beta\gamma}^{ijk} & = &
- {1 \over 3} {\rm FL}^* \left[
\frac{\partial^{2} G_{\alpha}}{\partial p_{k} \partial p_{l}}
p_{l \beta\gamma}^{ij}
+ \frac{\partial^{2} G_{\beta}}{\partial p_{k} \partial p_{l}}
p_{l \gamma\alpha}^{ij}
+ \frac{\partial^{2} G_{\gamma}}{\partial p_{k} \partial p_{l}}
p_{l \alpha\beta}^{ij}\right. 
+ \nonumber \\
& & + \frac{\partial^{3} G_{\alpha}}{\partial p_{k} \partial
p_{m}
\partial p_{n}} 
\left( p_{m\beta}^{i} p_{n\gamma}^{j} 
-  p_{m\gamma}^{i} p_{n\beta}^{j}\right) 
+ \frac{\partial^{3} G_{\beta}}{\partial p_{k} \partial
p_{m}
\partial p_{n}} 
\left( p_{m\gamma}^{i} p_{n\alpha}^{j} 
- p_{m\alpha}^{i} p_{n\gamma}^{j}\right) 
+ \nonumber \\
& & \left. + \frac{\partial^{3} G_{\gamma}}{\partial p_{k}
\partial p_{m} \partial p_{n}} 
\left( p_{m\alpha}^{i} p_{n\beta}^{j} 
- p_{m\beta}^{i} p_{n\alpha}^{j}\right) 
\right]
\label{3.52}
\end{eqnarray}
\end{subequations}

The functions ${\rm FL}^* p_{j\alpha}^{i}$ and ${\rm
FL}^* p_{k\alpha\beta}^{ij}$ are given by the expressions:

\begin{equation}
{\rm FL}^* p_{j\mu}^{i} = 
W_{jk} {\rm FL}^* \frac{\partial^{2} G_{\mu}}{\partial p_{k} \partial
p_{i}}
\label{3.53}
\end{equation}

\begin{eqnarray}
{\rm FL}^* p_{k\mu\nu}^{ij} & = & 
\frac{\partial W_{lm}}{\partial\dot{q}^{k}} 
{\rm FL}^* \left(
\frac{\partial^{2} G_{\mu}}{\partial p_{i} \partial p_{l}} 
\frac{\partial^{2} G_{\nu}}{\partial p_{m} \partial p_{j}} 
- \frac{\partial^{2} G_{\nu}}{\partial p_{i} \partial p_{l}} 
\frac{\partial^{2} G_{\mu}}{\partial p_{m} \partial p_{j}} 
\right) + \nonumber \\
& & + W_{lm}W_{kn} {\rm FL}^* \frac{\partial}{\partial p_{n}} 
\left(
\frac{\partial^{2} G_{\mu}}{\partial p_{i} \partial p_{l}} 
\frac{\partial^{2} G_{\nu}}{\partial p_{m} \partial p_{j}} 
- \frac{\partial^{2} G_{\nu}}{\partial p_{i} \partial p_{l}} 
\frac{\partial^{2} G_{\mu}}{\partial p_{m} \partial p_{j}} 
\right)
\label{3.54}
\end{eqnarray}

The above derivations illustrate how the lagrangian gauge structure
tensors can be derived from the BV field-antifield action (\ref{3.D}).

\end{document}